\documentclass[12pt]{article}
\usepackage{amssymb,epsfig,amsmath,amsfonts,graphicx}

\newcommand{\Tr}{\text{\,Tr\,}}
\newcommand{\eff}{\text{eff}}

\newcommand{\diag}{\text{\,diag\,}}
\newcommand{\LL}{\mathcal{L}}
\newcommand{\OO}{\mathcal{O}}
\newcommand{\TeV}{\text{ TeV }}
\newcommand{\GeV}{\text{ GeV}}

\newcommand{\half}{\frac{1}{2}}
\newcommand{\hc}{\text{ h.c. }}
\newcommand{\identity}{{\rlap{1} \hskip 1.6pt \hbox{1}}}
\newcommand{\lsim}{\,\raise.3ex\hbox{$<$\kern-.75em\lower1ex\hbox{$\sim$}}\,}
\newcommand{\gsim}{\,\raise.3ex\hbox{$>$\kern-.75em\lower1ex\hbox{$\sim$}}\,}

\begin{document}

\begin{titlepage}
\begin{flushright}
HUTP-03/A019\\
hep-ph/0303001\\
\end{flushright}
\vskip 2cm
\begin{center}
{\large\bf Little Higgs and Custodial $SU(2)$ }
\vskip 1cm
{\normalsize
Spencer Chang and Jay G. Wacker\\
\vskip 0.5cm

Jefferson Physical Laboratory\\
Harvard University\\
Cambridge, MA 02138\\
\vskip .1in}
\end{center}

\vskip .5cm

\begin{abstract}
In this note we present a little Higgs model that has custodial $SU(2)$
as an approximate symmetry.  This theory is a simple modification of the ``Minimal Moose''
with $SO(5)$ global symmetries protecting the Higgs mass.
This allows for a simple limit where
TeV physics makes small contributions to precision electroweak observables.
The spectrum of particles and their couplings to Standard Model fields are studied in detail.
At low energies this model has 
two Higgs doublets and it favours a light Higgs from precision
electroweak bounds, though for different reasons than in the Standard Model. 
The limit on the breaking scale, $f$, is roughly 700 GeV, with
a top partner of 2 TeV, $W'$ and $B'$ of 2.5 TeV, and 
heavy Higgs partners of 2 TeV.  These particles are easily accessible
at hadron colliders.
\end{abstract}

\end{titlepage}

\section{Introduction}
\setcounter{equation}{0}
\renewcommand{\theequation}{\thesection.\arabic{equation}}

Recently the little Higgs mechanism has been proposed as a way to stabilise 
the weak scale from the radiative corrections of the Standard Model. 
In little Higgs models the Standard Model Higgs boson is 
a pseudo-Goldstone and is kept light by approximate non-linear symmetries
\cite{Arkani-Hamed:2001nc,Arkani-Hamed:2002pa,Arkani-Hamed:2002qx,Gregoire:2002ra,Arkani-Hamed:2002qy,Low:2002ws,Kaplan:2003uc}, 
see \cite{Wacker:2002ar,Schmaltz:2002wx} for summaries of the physics
and \cite{Burdman:2002ns,Han:2003wu,Dib:2003zj,Han:2003gf} for more detailed phenomenology. 
The little Higgs mechanism requires that two separate couplings communicate to the Higgs 
sufficient breaking of the non-linear symmetry to generate a Higgs mass.
The weak scale is radiatively generated two loop factors beneath the cut-off
$\Lambda \sim 10 - 30$ TeV.  Little Higgs models predict a host
of new particles at the TeV scale that cancel the low energy
quadratic divergences to the Higgs mass from Standard Model
fields.   The little Higgs mechanism has particles of the 
\begin{it}same\end{it} spin cancel the quadratic divergences to the Higgs
mass, i.e. a fermion cancels a quadratic divergence from a fermion.  
In models described by ``theory space,'' such as the Minimal Moose, particles of the 
same spin and quantum numbers cancel quadratic divergences, for example a TeV scale
vector that transforms as a $SU(2)_L$ triplet cancels the $W$ 
quadratic divergence.  
To avoid fine-tuning the Higgs potential by
more then $\mathcal{O}(20\%)$ the top quark one loop
quadratic divergence should be cut off by roughly $2 \TeV$,  the quadratic
divergence from $SU(2)_L$  should be cut off by $5 \TeV$, while the quadratic
divergence from the Higgs quartic coupling should be cut off by $8\TeV$.

These TeV scale particles are heavier than the current experimental
limits on direct searches, however these particles may have effects
at low energy by contributing to higher dimension operators in the Standard
Model after integrating them out.
The effects of integrating out the TeV scale partners have been considered in
\cite{Chivukula:2002ww,Hewett:2002px,Csaki:2002qg}
and have provided constraints on some little Higgs
models from precision electroweak observables.   
Understanding what constraints are placed on each little Higgs model 
is a detailed question but their themes are the same throughout.  
The arguments for the most severe constraints on the ``littlest Higgs''
model discussed in \cite{Hewett:2002px,Csaki:2002qg} arise from the 
massive vector bosons interactions because they can contribute to low energy four 
Fermi operators and violate custodial $SU(2)$.  
Consider the $B'$ which cancels the quadratic divergence of the $B$,
the gauge eigenstates are related to the physical eigenstates by:
\begin{eqnarray}
B = \cos \theta' B_1  + \sin \theta' B_2
\hspace{0.5in}
B' = \cos \theta' B_2  - \sin \theta' B_1
\end{eqnarray}
where the mixing angles are related to the high energy gauge 
couplings through:
\begin{eqnarray}
g_1' = \frac{g'}{\cos \theta'}
\hspace{0.5in}
g_2' = \frac{g'}{\sin \theta'}
\end{eqnarray}
where $g'$ is the low energy $U(1)_Y$  gauge coupling.
With the Standard Model fermions charged only under $U(1)_1$, the
coupling to the $B'$ is:
\begin{eqnarray}
\LL_{B' \text{F Int}} = g' \tan \theta'\; B'_\mu \;j^\mu_{U(1)_Y}
\end{eqnarray}
where $j^\mu_{U(1)_Y}$ is the $U(1)_Y$ current.
The mass of the $B'$ goes as:
\begin{eqnarray}
m^2_{B'} \sim \frac{g'{}^2 f^2}{\sin^2 2\theta'}
\end{eqnarray}
where $f$ is the breaking scale.  After integrating out the $B'$
there is a four Fermi coupling of the form:
\begin{eqnarray}
\LL_{4 \text{ Fermi}} \sim  \frac{\sin^4 \theta'}{f^2} \big(j^\mu_{U(1)_Y}\big)^2
\end{eqnarray}
The coefficient of this operator needs to be roughly less than 
$(6 \TeV)^{-2}$ and can be achieved keeping $f$ fixed as 
$\theta' \rightarrow 0$.

 The little Higgs boson also couples to the $B'$ through
the current:
\begin{eqnarray}
\LL_{B' \text{ H Int}} \sim  g' \cot 2 \theta' 
 \;B'_\mu\; (i h^\dagger \overleftrightarrow{D}^\mu h).
\end{eqnarray}
Integrating out the $B'$ induces several dimension 6 operators  including:
\begin{eqnarray}
\LL_{ (h^\dagger D h)^2} \sim  \frac{\cos^2 2 \theta '}{f^2}\big(
(h^\dagger D h)^2 +\hc \big)
\end{eqnarray}
This operator violates custodial $SU(2)$ and after electroweak
symmetry breaking it lowers the mass of  the $Z^0$ and gives a positive 
contribution to the $T$ parameter.  This operator needs to be suppressed
by $(5 \TeV)^{-2}$.   Thus the Higgs coupling prefers the limit 
$\theta' \rightarrow \frac{\pi}{4}$. 
There are additional contributions to the $T$ parameter that can negate
this effect, this argument shows the potential tension in little
Higgs models that could push the limits on $f$ to 3 -- 5 TeV.  

The reason why the $B'$ contributes to an $SU(2)_C$ violating
operator is because it, like the $B$, couples
as the $T^3$ generator of $SU(2)_r$\footnote{
Recall that in the limit that $g' \rightarrow 0$ there
is an $SU(2)_l\times SU(2)_r$ symmetry of the Higgs and gauge sector.
Only the $T^3$ generator is gauged inside $SU(2)_r$ and $g'$ can
be viewed as a spurion parameterising the breaking.
After electroweak symmetry breaking $SU(2)_l\times SU(2)_r\rightarrow
SU(2)_C$.} 
and  its interactions explicitly break $SU(2)_C$.  The most straight-forward way of
softening this effect is to complete the $B'$ into a full triplet
of $SU(2)_C$\footnote{
The $W'$ transforms as a triplet of $SU(2)_C$ so no 
$SU(2)_C$ violating operators are generated by its interactions.
}.
This modification adds an additional charged
vector boson $W^{r\,\pm}$.  By integrating out these charged gauge
bosons there is another dimension 6 operator that gives a mass
to the $W^\pm$ compensating for the effect from the $B'$.  
This can be implemented by gauging $SU(2)_r$ instead of $U(1)_2$.
At the TeV scale $SU(2)_r \times U(1)_1 \rightarrow U(1)_Y$.
With these additional vector bosons, it is possible to take the
$\theta' \rightarrow 0$ limit without introducing large $SU(2)_C$ 
violating effects while simultaneously decoupling the Standard Model fermions
from the $B'$ and keeping the breaking scale $f$ fixed. 
Thus the limits on the model will roughly reduce to limits on
the $SU(2)_r$ coupling and the breaking scale.  

It is not necessary to have a gauged $SU(2)_r$ for the little Higgs
mechanism to be viable because the constraining physics is not 
crucial for stabilising the weak scale. 
The $B'$ is canceling the $U(1)_Y$ quadratic divergence that is only
borderline relevant for a cut-off $\Lambda \lsim 10 - 15$ TeV but
is providing some of the main limits through its interactions with
the Higgs and the light fermions.
The light fermions play no role in
the stability of the weak scale, therefore the limits from their
interactions can be changed without altering the little Higgs mechanism.
It is straightforward to avoid the strongest constraints \cite{Arkani:PC}.
The easiest possibility is to only gauge $U(1)_Y$ and accept its quadratic 
divergence with a  cut-off at 10 -- 15 TeV.
Another way of dealing with this issue
is to have the fermions charged equally under both $U(1)$ gauge groups.
With this charge assignment the fermions decouple from the $B'$ 
when $\theta' \rightarrow \frac{\pi}{4}$ which also decouples the
little Higgs from the $B'$.  There are other ways of decoupling
the $B'$ by mixing the Standard Model fermions with multi-TeV 
Dirac fermions in a similar fashion as \cite{Kaplan:2003uc}.
However having a gauged $SU(2)_r$ allows for a particularly 
transparent limit where TeV scale physics is parametrically safe and does 
not add significant complexity.  

In this note a new little Higgs model is presented that has the property 
that it has custodial $SU(2)$ as an approximate symmetry of the Higgs sector
by gauging $SU(2)_r$ at the TeV scale.
To construct a little Higgs theory with an $SU(2)_C$ symmetry we
can phrase the model building issue as: ``Find a little Higgs
theory that has the Higgs boson transforming as a $\mathbf{4}$ of $SO(4)$.''
This is precisely the same challenge as finding a little Higgs
theory that has a Higgs transforming as a $\mathbf{2_\half}$ of $SU(2)_L\times U(1)_Y$.  
In the latter case it was necessary to find a group that contained $SU(2)\times U(1)$ 
and where the adjoint  of the group 
had a field transforming as a $\mathbf{2_\half}$ and the simplest scenario
is $SU(3)$ where $\mathbf{8} \rightarrow \mathbf{3_0} + \mathbf{2_\half} + \mathbf{1_0}$.  
For a $\mathbf{4}$ of $SO(4)$ the simplest possibility is $SO(5)$ where an adjoint of $SO(5)$ 
decomposes into $\mathbf{10} \rightarrow \mathbf{6} + \mathbf{4}$.
The generators of $SO(5)$ are labeled
as $T^l$, $T^r$, and $T^v$ for the $SU(2)_l$, $SU(2)_r$ and
$SO(5)/SO(4)$ generators respectively.

The model presented in this paper is a slight variation of the ``Minimal Moose'' 
\cite{Arkani-Hamed:2002qx} that has four non-linear sigma model fields, $X_i$:
\begin{eqnarray}
X_i = \exp(i x_i/f)
\end{eqnarray}
where $x_i$ is the linearised field and $f$ is the breaking scale associated
with the non-linear sigma model. 
The Minimal Moose has an $[SU(3)]^8$ global symmetry associated with 
transformations on the fields:
\begin{eqnarray}
X_i \rightarrow L_i X_i R_i^\dagger
\end{eqnarray}
with $L_i, R_i \in SU(3)$.
To use the $SO(5)$ group theory replace the $SU(3) \rightarrow SO(5)$
keeping the ``Minimal Moose module'' of four links with an $[SO(5)]^8$.
The Minimal Moose had an $SU(3)\times [SU(2)\times U(1)]$ gauged
where the $[SU(2)\times U(1)]$ was embedded
inside $SU(3)$ while this model has an $SO(5)\times[SU(2)\times U(1)]$ 
gauge symmetry, using the $T^{l\,a}$ generators for $SU(2)$ and $T^{r\,3}$ 
generator for $U(1)$. 

The primary precision electroweak constraints arise from integrating
out the TeV scale vector bosons.  In this model there is a full
adjoint of $SO(5)$ vector bosons.    Under $SU(2)_l\times SU(2)_r$ they 
transform as:
\begin{eqnarray}
W^l \sim (\mathbf{3}_l, \mathbf{1}_r)
\hspace{0.15in}
W^r \sim ( \mathbf{1}_l, \mathbf{3}_r)
\hspace{0.15in}
V \sim (\mathbf{2}_l, \mathbf{2}_r)
\end{eqnarray}
Because only $U(1)_Y$ is gauged inside $SU(2)_r$ the $W^{r\,a}$
split into $W^{r\pm}$ and $W^{r3}$.  The $W^{r3}$ is the
mode that is responsible for canceling the one loop quadratic
divergence of the $U(1)_Y$ gauge boson and is denoted 
as the $B'$.  Finally the $V$ has the same quantum
numbers as the Higgs boson but has no relevant interactions to 
Standard Model fields.

In the limit where the $SO(5)$ gauge coupling becomes large
the Standard Model $W$ and $B$ gauge bosons become large
admixtures of the $SU(2)\times U(1)$ vector bosons.  
This means that the orthogonal combinations, the $W'$ and $B'$, 
are dominantly admixtures of the $SO(5)$ vector bosons.
The Standard Model fermions are charged only under $SU(2)\times U(1)$
which means that the TeV scale vector bosons decouple from the Standard Model
fermions in this limit.

In the remaining portion of the paper the explicit model is presented and
the spectrum is calculated along with the relevant couplings for precision electroweak
observables in Section \ref{Sec: Model}.  This model has two light Higgs doublets
with the charged Higgs boson being the heaviest of the physical Higgs states
because of the form of the quartic potential.  This potential is different than 
the quartic potential of the MSSM and has the property that it forces the Higgs
vacuum expectation values to be complex, breaking $SU(2)_C$ in the process.  This
will result in the largest constraint on the model.
In Section \ref{Sec: PEW} the TeV scale particles are integrated out and their effects
discussed in terms of the dimension 6 operators that are the
primary precision electroweak observables.   For an $SO(5)$ coupling of $g_5\sim 3$ 
and $f\sim 700 \GeV$ and for $\tan \beta\lsim 0.3$ the model has no constraints 
placed on it.  The limit on $\tan \beta$ ensures a light Higgs with mass
in the 100 -- 200 GeV range.
With the rough limits on the parameters, the masses for the
relevant TeV scale fields are roughly $2.5 \TeV$ for the gauge bosons, $2 \TeV$ for the
top partner, and $2\TeV$ for the Higgs partners.
Finally in Section \ref{Sec: Conclusions} the outlook for this model 
and the state of little Higgs models in general is discussed.

\section{$SO(5)$ Minimal Moose}
\setcounter{equation}{0}
\renewcommand{\theequation}{\thesection.\arabic{equation}}
\label{Sec: Model}

Little Higgs models are theories of electroweak symmetry breaking
where the Higgs is a pseudo-Goldstone boson and can be described
as gauged non-linear sigma models.  In this model there 
is an $SO(5)\times[SU(2)\times U(1)]$
gauge symmetry with standard gauge kinetic
terms with couplings $g_5$ and $g_2, g_1$, respectively.  
There are four non-linear sigma model fields, $X_i$, that
transform under the global $[SO(5)]^8= [SO(5)_L]^4\times [SO(5)_R]^4$ as:
\begin{eqnarray}
X_i \rightarrow  L_i X_i R_i^\dagger .
\end{eqnarray}
Under a gauge transformation the non-linear sigma model fields
transform as:
\begin{eqnarray}
X_i \rightarrow  G_{2,1} X_i G_5^\dagger
\end{eqnarray}
where $G_5$ is an $SO(5)$ gauge transformation and $G_{2,1}$ is an $SU(2)\times
U(1)$ gauge transformation with $SU(2)\times U(1)$ embedded inside 
$SO(4) \simeq SU(2)_l \times SU(2)_r$, see Appendix \ref{App: Generators}
for a summary of the conventions.  
The gauge symmetries explicitly break the global $[SO(5)]^8$ symmetry and
the gauge couplings $g_5$ and $g_{2,1}$ can be viewed as spurions.
Notice that $g_5$ only breaks the $[SO(5)_R]^4$ symmetry, while
$g_{2,1}$ only breaks the $[SO(5)_L]^4$ symmetry.

The non-linear sigma model fields,
$X_i$, can be written in terms of linearised fluctuations around
a vacuum $\langle X_i \rangle = \identity$:
\begin{eqnarray}
X_i = \exp( i x_i/f)
\end{eqnarray}
where $f$ is the breaking scale of the non-linear sigma model and $x_i$
are adjoints under the diagonal global $SO(5)$.    The interactions of
the non-linear sigma model become strongly coupled at roughly $\Lambda \simeq 4\pi f$
where new physics must arise.
The kinetic term for the non-linear sigma model fields is:
\begin{eqnarray}
\LL_{\text{nl$\sigma$m Kin}} = \half \sum_i f^2 
\Tr D_\mu X_i D^\mu X_i^\dagger .
\end{eqnarray}
where the covariant derivative is:
\begin{eqnarray}
D_\mu X_i &=& \partial_\mu X_i 
-i g_5 X_i T^{[mn]} W^{[mn]}_{SO(5)}{}_\mu
+ i  \big( g_2\, T^{la}\, W^{la}_\mu 
+ g_1\, T^{r3}\, W^{r3}_\mu\big)X_i
\end{eqnarray}
where $W^{[mn]}_{SO(5)}$ are the $SO(5)$ gauge bosons, $W^{la}$ are 
the $SU(2)$ gauge bosons and $W^{r3}$ is the $U(1)$ gauge boson.
One linear combination of linearised fluctuations is eaten:
\begin{eqnarray}
\rho \propto x_1 + x_2 + x_3 + x_4 
\end{eqnarray}
leaving three physical pseudo-Goldstone bosons in adjoints of the global $SO(5)$ that
decompose under $SU(2)_l\times SU(2)_r$ as:
\begin{eqnarray}
\phi^l \sim ( \mathbf{3}_l, \mathbf{1}_r)
\hspace{0.3in}
\phi^r \sim ( \mathbf{1}_l, \mathbf{3}_r)
\hspace{0.3in}
h \sim ( \mathbf{2}_l, \mathbf{2}_r)
\end{eqnarray}
Under $U(1)_Y$, $\phi^r$ splits into $\phi^{r\,0}$ and $\phi^{r\,\pm}$.

\subsubsection*{Radiative Corrections}

There are no one loop quadratic divergences to the masses of
the pseudo-Goldstone bosons from the gauge sector because all the 
non-linear sigma model fields are bi-fundamentals of the gauge groups.  
This occurs because the $g_5$ gauge couplings break only the $SO(5)_{R_i}$ global
symmetries, while the $g_{2,1}$ couplings only break the $SO(5)_{L_i}$
symmetries.  
To generate a mass term it must arise from an operator $|\Tr X_i X_j^\dagger|^2$
and needs to simultaneously break both the left and right global symmetries.
This requires both the $g_5$ and $g_{2,1}$ gauge couplings which cannot 
appear as a quadratic divergence until two loops.
This can be verified with the Coleman-Weinberg potential 
\cite{Coleman:jx}.  
In this case the mass squared matrix
is:
\begin{eqnarray}
\left(\begin{array}{cc} W^A_{5}& W^{A'}_{2,1}\end{array}\right)
\left(
\begin{array}{cc}
g_5^2 f^2 \Tr T^A X_i X_i^\dagger T^B&
g_5 g_{2,1} f^2 \Tr T^A X_i T^{B'} X_i^\dagger \\
g_5 g_{2,1} f^2 \Tr T^{A'} X_i^\dagger T^B X_i &
g_{2,1}^2 f^2 \Tr T^{A'} X_i^\dagger X_i T^{B'}
\end{array}
\right)
\left(\begin{array}{c} W^B_{5}\\ W^{B'}_{2,1}\end{array}\right)
\end{eqnarray}
Because the fields are unitary matrices, the entries along the
diagonal are independent of the background field, $x_i$, and so is
the trace of the mass squared.  Therefore:
\begin{eqnarray}
V_{\text{1 loop CW $\Lambda^2$}} = \frac{3}{32 \pi^2} \Lambda^2 \Tr M^2[x_i]
= \text{Constant}
\end{eqnarray}
There are one loop logarithmically
divergent, one loop finite and two loop quadratic divergences from
the gauge sector.  All these contributions result 
in masses for the pseudo-Goldstone bosons that are parametrically two
loop factors down from the cut-off and are $\mathcal{O}(g^2 f/4\pi)$ 
in size.  

\subsection{Vector Bosons: Masses and Couplings}

The masses for the vector bosons arise as the lowest order expansion
of the kinetic terms for the non-linear sigma model fields.
The  $SO(5)$ and $SU(2)$ $W^l$  vector bosons mix as do the $SO(5)$ and $U(1)$
$W^{r\,3}$ vector bosons.  They can be diagonalised with the following
transformations:
\begin{eqnarray}
\nonumber
B &=& \cos\theta' W^{r3} - \sin\theta' W_{SO(5)}^{r3}
\hspace{0.5in}
B'= W'{}^{\,r3} = \sin \theta' W^{r3} + \cos \theta'  W_{SO(5)}^{r3}\\
\nonumber
W^a &=& \cos\theta W^{la} - \sin\theta W_{SO(5)}^{la}
\hspace{0.5in}
W'{}^a=W'{}^{\,la} = \sin \theta W^{la} + \cos \theta W_{SO(5)}^{la}
\end{eqnarray}
where the mixing angles are related to the couplings by:
\begin{eqnarray}
\nonumber
\cos \theta' &=& g'/g_1 \hspace{0.5in} \sin \theta' = g'/g_5 \\
\cos \theta &=& g/g_2 \hspace{0.5in} \sin \theta = g/g_5 
\end{eqnarray}
The angles $\theta$ and $\theta'$ are not independent and are 
related through the weak mixing angle by:
\begin{eqnarray}
\tan \theta_{\text{w}} = \frac{\sin \theta'}{\sin \theta}
\end{eqnarray}
and since $\theta_\text{w} \simeq 30^\circ$, 
$\sin \theta \simeq \sqrt{3} \sin \theta'$.

The masses for the vectors can be written in terms of the electroweak gauge
couplings and mixing angles:
\begin{eqnarray}
m^2{}_{W'}  =  \frac{16 g^2 f^2}{\sin^2 2\theta} 
\hspace{0.5in}
m^2{}_{B'} =  \frac{16 g'{}^2 f^2}{\sin^2 2 \theta'} 
\hspace{0.5in}
m^2{}_{W^{r\,\pm}} = \frac{16 g'^2 f^2}{\sin^2 2 \theta'} \cos^2 \theta'
\end{eqnarray}
These can be approximated in the $\theta'\rightarrow 0$ limit as:
\begin{eqnarray}
m^2_{B'} \simeq m^2_{W'} ( 1 - \frac{2}{3} \sin^2 \theta)
\hspace{0.5in}
m^2_{W^{r\,\pm}} \simeq m^2_{W'} ( 1 -  \sin^2 \theta)
\end{eqnarray}
Note that the $B'$, the mode that is canceling the quadratic divergence of 
the $B$, is not anomalously light\footnote{ 
The $B'$ in the ``littlest Higgs'' is 
a factor of $\sqrt{5}$ lighter and in the $SU(3)$ Minimal Moose 
it is a factor of $\sqrt{3}$ lighter.}.  
The $U(1)_Y$ quadratic divergence 
is borderline relevant for naturalness and could be neglected if the 
cut-off $\Lambda \lsim 10 - 15 \TeV$.   The corresponding mode is
contributing to electroweak constraints but doing 
little to stabilise the weak scale quantitatively.

The Higgs boson couples to these vector bosons through the currents: 
\begin{eqnarray}
\nonumber
j^{\mu\;a}_{W'} & = &  g \cot 2\theta j_{\text{H}}^{\mu\;a}
= \frac{g \cos{2\theta}}{2 \sin 2\theta} 
\;i h^\dag \sigma^a \overleftrightarrow{D}^\mu h\\
j^{\mu}_{B'} & = &g' \cot 2 \theta' j_{\text{H}}^{\mu}
= -\frac{g' \cos{2\theta'}}{2 \sin 2\theta'} 
\;i h^\dagger \overleftrightarrow{D}^\mu h
\label{Eq: Higgs Currents}
\end{eqnarray}
where $D_\mu$ is the Standard Model covariant derivative and
$j_{\text{H}}^{\mu\;a}$ is the $SU(2)_L$ current that the Higgs
couples to and $j_{\text{H}}^{\mu}$ for $U(1)_Y$.

The Higgs also couples to the charged $SU(2)_r$ vector bosons through:
\begin{eqnarray}
\nonumber
j^{\mu}_{W^{r\,+}} & = & -\frac{g' \cos\theta'}{\sqrt{2}\sin{2\theta'}} 
\;i h  D^\mu h \\
j^{\mu}_{W^{r\,-}} &=& j^{\mu}_{W^{r\,+}}{}^\dagger .
\end{eqnarray}
where the $SU(2)_L$ indices are contracted with the alternating tensor.
Notice that this interaction is not invariant under rephasing of the Higgs:
$h\rightarrow e^{i\phi} h$ sends $j_{W^{r\,+}} \rightarrow e^{2 i\phi} j_{W^{r\,+}}$.

\subsection{Scalar Masses and Interactions}

In order to have viable electroweak symmetry breaking there must
be a significant quartic potential amongst the light fields.
It is useful to define the operators:
\begin{eqnarray}
\mathcal{W}_i = X_i X_{i+1}^\dagger X_{i+2} X_{i+3}^\dagger
\end{eqnarray}
where addition in $i$ is modulo 4.
There is a potential for the non-linear sigma model fields:
\begin{eqnarray}
\label{Eq: Pot}
\hspace{-0.4in} 
\LL_{\text{Pot.}} &=&
\lambda_1 f^4 \Tr  \mathcal{W}_1
+ \lambda_2 f^4 \Tr  \mathcal{W}_2 +\hc
\end{eqnarray}
There is a $\mathbb{Z}_4$ symmetry where the link fields cycle as $X_i \to X_{i+j}$ 
that forces $\lambda_1 = \lambda_2$.
This is an approximate symmetry that is kept to $\mathcal{O}(10\%)$.
This potential gives a mass to one linear combination of linearised
fields:
\begin{eqnarray}
u_H = \frac{1}{2}(x_1 - x_2 + x_3 - x_4).
\end{eqnarray}
The other two physical modes are the little Higgs and are classically
massless:
\begin{eqnarray}
u_1 = \frac{1}{\sqrt{2}}(x_1 - x_3) \hspace{0.6in} 
u_2 = \frac{1}{\sqrt{2}}(x_2 - x_4).
\end{eqnarray}
The potential in Eq. \ref{Eq: Pot} can be expanded out in terms of these 
physical eigenmodes using the Baker-Campbell-Hausdorff formula:
\begin{eqnarray}
\nonumber
\hspace{-0.3in}
\LL_{\text{Pot.}} &=&
\lambda_1 f^4 \Tr 
\exp\Big(  2 i \frac{u_H}{f} + \half \frac{[u_1,u_2]}{f^2} + \cdots\Big)\\
&&+ \lambda_2 f^4 \Tr 
\exp\Big( - 2 i \frac{u_H}{f} + \half \frac{[u_1,u_2]}{f^2} + \cdots\Big) +\hc
\label{Eq: Pot2}
\end{eqnarray}
The low energy quartic coupling is related to the previous couplings through:
\begin{eqnarray}
\nonumber
\lambda^{-1} &=& \lambda_1^{-1} + \lambda_2^{-1}
\hspace{0.4in}
\lambda_1 = \lambda/ \cos^2 \vartheta_\lambda
\hspace{0.3in}
\lambda_2 = \lambda/ \sin^2 \vartheta_\lambda
\end{eqnarray}
The approximate $\mathbb{Z}_4$ symmetry sets 
$\theta_\lambda \approx \frac{\pi}{4}$ and the
symmetry breaking parameter is 
$\cos 2 \vartheta_\lambda \sim \mathcal{O}(10^{-1})$.
The mass of the heavy scalar is:
\begin{eqnarray}
m^2_{u_H} = \frac{16 \lambda f^2}{\sin^2 2 \vartheta_\lambda} .
\end{eqnarray}
After integrating out the massive mode the resulting potential for the 
little Higgs is the typical commutator potential: 
\begin{eqnarray}
V(u_1,u_2) &=&  - \lambda \Tr [ u_1, u_2]^2 +  \cdots
\label{Eq: Pot3}
\end{eqnarray}

In order to have stable electroweak symmetry breaking it is
necessary to have a mass term $i h_1^\dagger h_2 + \hc$.
This can arise from a potential of the form:
\begin{eqnarray}
\LL_{\text{$T^{r\,3}$ Pot.}} &=&
i \epsilon f^4 \Tr  T^{r\,3}\,\big( 
\mathcal{W}_1 + \mathcal{W}_2 + \mathcal{W}_3 + \mathcal{W}_4
\big)
+\hc
\end{eqnarray}
where $T^{r\,3}$ is the  $U(1)$ generator.
The size of the effects are radiatively stable and they are set to be a loop factor less 
than $\lambda$, $\epsilon \sim 10^{-2} \lambda$.
The coefficients are taken to be pure imaginary because the imaginary coefficient
will be necessary to ensure stable electroweak symmetry breaking while
the real parts are small $SO(5)$ splittings amongst the various modes.
Expanding this out to quadratic order:
\begin{eqnarray}
V_{\text{$T^{r3}$ Pot.}} &=& 
4 \epsilon f^2 \Tr T^{r3} i [u_1,u_2]  +\cdots
\end{eqnarray}
In terms of the Higgs doublets, $h_{1,2} \in u_{1,2}$, the potentials are:
\begin{eqnarray}
V(h_1, h_2) \simeq \frac{\lambda}{2}
\big( |h_1^\dagger h_2 - h_1^\dagger h_2|^2 + 4 |h_1 h_2|^2  \big)
+ (4i \epsilon f^2 h_1^\dagger h_2 +\hc)
\end{eqnarray}
where the $h_1 h_2$ term is contracted with the $SU(2)$ alternating tensor.
This potential is not the same as the MSSM potential and will
lead to a different Higgs sector\footnote{
In the $SU(3)$ Minimal Moose the Higgs potential was identical to
the the MSSM because of the close relation between little Higgs
theories and orbifolded extra dimensions, see \cite{Gregoire:2002ra} for
the precise relation.}.   
There are radiative corrections to this potential whose largest effect
gives soft masses of $\mathcal{O}(100\GeV)$ to the doublets:
\begin{eqnarray}
\nonumber
V_{\text{eff}} &\simeq& \frac{\lambda}{2}
\big( |h_1^\dagger h_2 - h_2^\dagger h_1|^2 + 4 |h_1 h_2|^2  \big)\\
&& \hspace{0.3in}+ \big((i b + m_{12}^2)  h_1^\dagger h_2 +\hc\big)
+ m_1^2 |h_1|^2 + m_2^2 |h_2|^2
\end{eqnarray}
where $b \approx 4 \epsilon f^2$.  Typically $m_{12}^2$ is taken to
be small to simplify the phenomenology so that the Higgs states
fall into CP eigenstates.   

\subsubsection*{Radiative Corrections}

There are no one loop quadratic divergences to the Higgs mass 
from the scalar potential\footnote{
More generally potentials that only contain any non-linear sigma model field
at most once can only give a quadratically divergent contribution to themselves.
}.
The symmetry breaking pattern in the potential is more
difficult to see, but notice that if either $\lambda_1$
or $\lambda_2$ vanished then there is a non-linear  symmetry
acting on  the fields:
\begin{eqnarray}
\nonumber
&&\delta_{\epsilon_1} u_1 = \epsilon_1+\cdots
\hspace{0.3in}
\delta_{\epsilon_1} u_2 = \epsilon_1+\cdots
\hspace{0.3in}
\delta_{\epsilon_1} u_H = -\frac{i}{4f} [\epsilon_1,u_1-u_2] +\cdots\\
&&\delta_{\epsilon_2} u_1 = \epsilon_2+\cdots
\hspace{0.3in}
\delta_{\epsilon_2} u_2 = \epsilon_2+\cdots
\hspace{0.3in}
\delta_{\epsilon_2} u_H = +\frac{i}{4f}[\epsilon_2,u_1-u_2] +\cdots .
\end{eqnarray}
$\Tr \mathcal{W}_1$  preserves the first non-linear symmetry but breaks the
second, while $\Tr \mathcal{W}_2$ preserves the second but breaks the first. 
Either symmetry is sufficient to keep $u_1$ and $u_2$ as exact Goldstones,
this is why $\lambda \rightarrow 0$ as $\lambda_1$ 
\begin{it}or\end{it} $\lambda_2 \rightarrow 0$.

There are one loop logarithmically divergent contributions to the masses of the little
Higgs as well as one loop finite and two loop quadratic divergences.
These are all positive and parametrically give masses of the order
of $\lambda^2 f/4 \pi$.

\subsection{Electroweak Symmetry Breaking}

At this point electroweak symmetry can be broken.
The little Higgs are classically massless but pick up 
$\mathcal{O}(100 \text{ GeV})$ masses from radiative corrections
to the tree-level Lagrangian.  The gauge and scalar corrections
to the little Higgs masses give positive contributions to the
mass squared of the little Higgs while fermions give negative 
contributions.    The mass matrix for the Higgs sector is of the form:
\begin{eqnarray}
\LL_{\text{Soft Mass}} = \left(\begin{array}{cc} h_1^\dagger & h_2^\dagger\end{array}\right)  
\left(\begin{array}{cc} m^2_1 & \mu^2 \\ \mu^*{}^2 & m^2_2\end{array}\right)
\left(\begin{array}{c} h_1 \\h_2\end{array}\right)
\end{eqnarray}
where $\mu^2 = m_{12}^2 + i b$.
To have viable electroweak symmetry breaking requires:
\begin{eqnarray}
\nonumber
m_1^2 > 0 &\hspace{0.2in}&
m_2^2 >0\\
\nonumber
m_1^2 m_2^2 - m_{12}^4 &>& 0\\
m_1^2 m_2^2 - m_{12}^4 - b^2 &<& 0 .
\end{eqnarray}
The vacuum expectation values are:
\begin{eqnarray}
\langle h_1 \rangle = \frac{1}{\sqrt{2}}\left( \begin{array}{c} 0\\ v \cos \beta \end{array}\right)
\hspace{0.3in}
\langle h_2 \rangle = \frac{1}{\sqrt{2}}\left( \begin{array}{c} 0\\ v\sin \beta e^{i\phi}\end{array}\right)
\end{eqnarray}
The potential has a flat direction when $\beta =0,\frac{\pi}{2}$ and when
$\phi = 0$.  Unfortunately when $\phi \ne 0$ custodial $SU(2)$ is broken\footnote{
This can be seen by going back to the $SO(4)$ description.  By having
a phase it is the same as having two $SO(4)$ vectors acquire vacuum expectation values 
in different directions leaving only $SO(2)\simeq U(1)_Y$ unbroken.}.  The
phase can be solved for in terms of the soft masses as:
\begin{eqnarray}
\cos\phi = \frac{m_{12}^2}{m_1 m_2} .
\end{eqnarray}
The breaking of $SU(2)_C$ by the Higgs sector provides one of the strongest 
limits on the model.  For simplicity $\mu^2 = i b$ is taken to be pure 
imaginary forcing $\phi = \frac{\pi}{2}$. Taking $\phi = \frac{\pi}{2}$ 
is clearly the worst-case scenario for $SU(2)_C$ and not generic because there is no reason 
for $m_{12}$ to be significantly smaller than any of the other masses.

The parameters of electroweak symmetry breaking can be solved for
readily in the limit $\phi = \frac{\pi}{2}$
in terms of the masses: 
\begin{eqnarray}
\nonumber
2 \lambda v^2 &=& ( m_1^2 + m_2^2)  \left( \frac{|b|}{m_1 m_2 } -1 \right)\\
\nonumber
\tan \beta &=& \frac{m_1}{m_2}\\
\tan 2 \alpha &=& \left(1-\frac{2m_1 m_2}{|b|}\right)\tan 2 \beta.
\end{eqnarray}
where $\alpha$ is the mixing angle for the $h^0 - H^0$ sector.
The  soft masses should not be much larger than $v$ otherwise it either
requires some tuning of the parameters so that $b \simeq m_1 m_2$ or $\lambda$ 
becoming large.  These arguments will change when $m_{12}^2 \ne 0.$  
The masses for the five physical Higgs are:
\begin{eqnarray}
\nonumber
m^2_{A^0} &=& m_1^2 + m_2^2\\
\nonumber
m^2_{H^\pm} &=& m_1^2 + m_2^2 + 2 \lambda v^2
\hspace{0.70in}=\hspace{0.25in} x \; m^2_{A^0}\\ 
\nonumber
m^2_{h^0} &=&  m^2_{H^\pm}\frac{\left( 1 - \sqrt{1 - m^2_0/m^2_{H^\pm}} \right)}{2}\\
m^2_{H^0} &=& 
m^2_{H^\pm}\frac{\left( 1 + \sqrt{1 - m^2_0/m^2_{H^\pm}} \right)}{2}
= m^2_{H^\pm} - m^2_{h^0}
\end{eqnarray}
where 
\begin{eqnarray}
x = |b|/m_1 m_2
\hspace{0.5in} m_0^2 = \frac{8\lambda v^2 \sin^2 2\beta}{x} 
\end{eqnarray}
The heaviest Higgs is the charged $H^\pm$ 
and this has consequences for precision electroweak observables.
The mass of the lightest Higgs is bounded by:
\begin{eqnarray}
\frac{1}{4}m^2_0 \le m^2_{h^0}\le \frac{1}{2}m^2_0
\end{eqnarray}
where the lower bound is saturated as $m^2_{H^\pm} \rightarrow \infty$ 
and the upper bound is saturated as $m^2_{H^\pm} \rightarrow m^2_0$.

\subsection{Fermions}

The Standard Model fermions are charged only under the
$SU(2)\times U(1)$ gauge group.  Since all the fermions
except the top quark couple extremely weakly to the
Higgs sector, the standard Yukawa coupling to the linearised Higgs doublets 
can be used without destabilising the weak scale.  These small
Yukawa couplings are spurions that simultaneously break flavour symmetries 
as well as the chiral symmetries of the non-linear sigma model.
There are many ways to covariantise these couplings but they only differ
by irrelevant operators.
\begin{eqnarray}
\LL_{\text{Yuk}} = y_u\, q h u^c 
+ y_d\, q h^\dagger d^c + y_e\, l h^\dagger e^c 
\end{eqnarray}
There is no symmetry principle that prefers type I or type II models.
This can have significant implications for Higgs searches.

The couplings of the Standard Model fermions to the heavy gauge bosons is:
\begin{eqnarray}
\label{Eq: Fermion Currents}
\LL_{\text{Int}} = g \tan \theta \; W'{}^a_\mu \, j_{\text{F}}^\mu{}_a
+ g'\tan \theta'\; B'_\mu\, j_{\text{F}}^\mu
\end{eqnarray}
where $j_{\text{F}}^{\mu\,a}$ is the $SU(2)_L$ electroweak current 
involving the Standard Model fermions and  
$j_{\text{F}}^\mu$ is the $U(1)_Y$ electroweak current
involving the Standard Model fermions.  In the limit $g_5 \rightarrow 
\infty$ both $\theta, \theta'\rightarrow 0$ and the TeV scale gauge bosons
decouple from the Standard Model fermions.

\subsection*{Top Yukawa}

The top quark couples strongly to the Higgs and how
the top Yukawa is generated is crucial for stabilising the weak scale.  
The top sector must preserve some
of the $[SO(5)]^8$ global symmetry that protects the Higgs mass.  There
are many ways of doing this but generically the mechanisms involve
adding additional Dirac fermions.  
To couple the non-linear sigma model fields to the quark doublets
it is necessary to
transform the bi-vector representation to the
bi-spinor representation, see Appendix \ref{App: Generators}.
The linearised fields are re-expressed as:
\begin{eqnarray}
\tilde{x}_i{}_\alpha{}^\beta = x_i{}_{[mn]}\sigma^{[mn]}{}_\alpha{}^\beta
\end{eqnarray}
where $m,n$ are $SO(5)$ vector indices running from 1 to 5, 
$\alpha,\beta$ are $SO(5)$ spinor indices running from 1 to 4 and $
\sigma^{[mn]}{}_\alpha{}^\beta$ are generators of $SO(5)$ in the spinor representation.
The exponentiated field, $\tilde{X}_i = \exp(i \tilde{x}_i/f)$, 
has well-defined transformation properties
under the global $SO(5)$'s and the
operator, $\mathcal{X} = (\tilde{X}_1 \tilde{X}_3^\dagger)$,
transforms only under the $SU(2)\times U(1)$ gauge symmetry:
\begin{eqnarray}
\mathcal{X} &\rightarrow& \tilde{G}_{2,1} \mathcal{X} \tilde{G}_{2,1}^\dagger
\end{eqnarray}
where $\tilde{G}_{2,1}$ is an $[SU(2)\times U(1)]\subset SO(5)$ 
gauge transformation in the spinor representation of $SO(5)$.

It is necessary to preserve some of the global $SO(5)$ symmetry 
in order to remove the one loop quadratic 
divergence to the Higgs mass from the top.   
As in the Minimal Moose, it is necessary to add additional fermions to fill 
out a full representation, in this case a $\mathbf{4}$ of $SO(5)$
for either the $q_3$ or the $u^c_3$.  The large top coupling is a result of 
mixing with this TeV scale fermion.
The most minimal approach is to complete the $q_3$ into:
\begin{eqnarray}
\mathcal{Q} = (q_3, \tilde{u}, \tilde{d})
\hspace{0.3in}
\mathcal{U}^c = ( 0_2, u^c_3, 0)
\end{eqnarray}
where $\tilde{u} \sim (\mathbf{3}_c, \mathbf{1}_{+\frac{2}{3}})$
and $\tilde{d} \sim (\mathbf{3}_c, \mathbf{1}_{-\frac{1}{3}})$
with charge conjugate fields $\tilde{u}^c$ and $\tilde{d}^c$
canceling the anomalies.  The top Yukawa coupling is generated by:
\begin{eqnarray}
\LL_{\text{top}} =
y_1 f \mathcal{U}^c \mathcal{X} \mathcal{Q}
+ y_2 f \tilde{u} \tilde{u}^c + \tilde{y}_2 f\tilde{d} \tilde{d}^c +\hc
\end{eqnarray}
The $\tilde{u}$ and $u_3^c$ mix with an angle $\vartheta_y$ 
and after integrating out the massive combination the low energy top Yukawa is given by:
\begin{eqnarray}
y_{\text{top}}^{-2} = 2 (|y_1|^{-2} + |y_2|^{-2} )
\hspace{0.6in}
\tan \vartheta_y = \frac{|y_1|}{|y_2|}.
\end{eqnarray}
After electroweak symmetry breaking the top quark and the top partner
pick up a mass:
\begin{eqnarray}
\label{Eq: Top Mass}
m_t = \frac{y_{\text{top}} v \cos \beta}{\sqrt{2} }
\hspace{0.5in}
m_{t'} = \frac{2\sqrt{2} y_{\text{top}} f}{\sin 2\vartheta_y}
\left( 1 - \frac{ v^2 \cos^2 \beta \sin^2 2 \vartheta_y}{32 f^2}\right) .
\end{eqnarray}
The decoupling limit is the $y_2 \rightarrow \infty$ limit where $\vartheta_y \rightarrow 0$.

\subsubsection*{Radiative Corrections}

The top coupling respects a global $SO(5)$ symmetry.
This ensures that there are no one loop quadratically divergent 
contributions to the Higgs mass  
and can be seen through the  Coleman-Weinberg potential.  
The one loop quadratic divergence is proportional to $\Tr M M^\dagger$, 
where $M \sim P_{\mathcal{U}^c} \mathcal{X}$ is the mass matrix
for the top sector in the background of the little Higgs  and 
$P_{\mathcal{U}^c} = \text{diag}(0,0,1,0)$ is a projection matrix 
from the $\mathcal{U}^c$.
Expanding this out:
\begin{eqnarray}
\nonumber
V_{\text{1 loop CW }\Lambda^2} &=& 
- \frac{12\Lambda^2}{32 \pi^2}\Tr P_{\mathcal{U}^c} 
\mathcal{X}{\cal X}^\dag P_{\mathcal{U}^c}\\ 
&\sim& \Tr P_{\mathcal{U}^c} = \text{Constant} 
\end{eqnarray}
which gives no one loop quadratic divergences to any of the $x_i$ fields.  
One loop logarithmically divergent, 
one loop finite and two loop quadratically divergent masses are generated
at the order $\OO(y^2_{\text{top}}f/4\pi)$.
Since the top only couples to $h_1$ amongst the light fields, it only generates a negative
contribution to $m_1^2$. This drives $\tan \beta$ to be small since this is the
only interaction that breaks the $h_1 \leftrightarrow h_2$ symmetry explicitly.

Note that the $\tilde{d}$ can be decoupled without affecting naturalness.  
This is because there is an accidental
$SU(3)$ symmetry that is identical to the $SU(3)$ symmetry of
the Minimal Moose.  
\begin{eqnarray}
\LL_{\text{Top}} = y_1 f u^c \tilde{u} +\frac{i}{\sqrt{2}} y_1 u^c h_1 q 
- \frac{1}{4} \frac{y_1}{f}  u^c h_1^\dagger h_1 \tilde{u} +
\cdots
\end{eqnarray}
is invariant under:
\begin{eqnarray}
\delta h_1 = \epsilon 
\hspace{0.3in}
\delta q = \frac{i\sqrt{2}}{f} \epsilon^* \tilde{u} 
\hspace{0.3in}
\delta\tilde{u} = \frac{i\sqrt{2}}{f} \epsilon q .
\end{eqnarray}
This can be seen by imagining an $SU(4)$ symmetry
acting on $\mathcal{X}$.  With only the $\tilde{u}$ there is an
$SU(3)$ acting in the upper components.  The $SU(4)$ symmetry
is just the $SO(6) \supset SO(5).$  The $SU(3)$ is not exact
but to quadratic order in $h$ it is an accidental symmetry.
This means that in principle it is possible to send 
$\tilde{y}_2 \rightarrow 4 \pi$ without affecting naturalness and
therefore it is safe to ignore this field.  
Performing the same calculation as above, the charged 
singlet, $\phi_1^{r\,\pm}$, gets a quadratically divergent mass
and is lifted to the TeV scale.  

\section{Precision Electroweak Observables}
\setcounter{equation}{0}
\renewcommand{\theequation}{\thesection.\arabic{equation}}
\label{Sec: PEW}

Throughout this note the scalings of the contributions of
TeV scale physics to precision electroweak observables have been 
discussed.  
The contributions to the higher dimension operators of the Standard Model
are calculated in this section.  The most physically transparent
way of doing this is to integrate out the heavy fields and then
run the operators down to the weak scale.   The most difficult
contribution to calculate is the custodial $SU(2)$ violating
operator because there are several sources.  Beyond that
there are four Fermi operators and corrections to the $Z^0$ and
$W^\pm$ interactions.  There are no important contributions to
the $S$ parameter besides the contributions from the Higgs
that turn out to be small.  In Sec. \ref{Sec: Limits} we
summarise the constraints on the model from precision electroweak
observables and state the limits on the masses.

\subsection{Custodial $SU(2)$}

Custodial $SU(2)$ provides limits on beyond the Standard Model physics.
When written in terms of the electroweak chiral Lagrangian, violations 
of $SU(2)_C$ are related to the operator:
\begin{eqnarray}
\OO_4 = c_4 v^2 \big(\Tr T_3 \omega^\dagger D_\mu \omega\big)^2
\end{eqnarray}
where $\omega$ are the Goldstone bosons associated with electroweak symmetry breaking.  
The coefficient of this operator is calculated in this section.  
This is directly related to $\delta \rho$.  
However, typically limits are stated in terms of the $T$ parameter
which is related to $\delta \rho_*$ which differs from $\delta\rho$ 
when there are modifications to the $W^\pm$ and $Z^0$ interactions
with Standard Model fermions.  In Sec \ref{Sec: Limits} this difference
is accounted for.

There are typically five new sources of custodial $SU(2)$ violation
in little Higgs models.  The first is from the non-linear sigma model
structure itself.  By expanding  the kinetic terms to quartic order
there are operators that give the $W^\pm$ and $Z^0$ masses.  
If $SU(2)_C$ had not been  broken by the vacuum expectation values of
the Higgs, then there could not be any operators that violate $SU(2)_C$.  
Custodial $SU(2)$ is only broken with the combination of the two
vacuum expectation values which means that the only possible operator that
could violate $SU(2)_C$ must be of the form $ (h_2^\dagger D h_1)^2$.  However, 
the kinetic terms for the non-linear sigma model fields never contain $h_1$ and
$h_2$ simultaneously meaning that any operator of this form is not present.  

\subsubsection*{Vector Bosons}

The second source of custodial $SU(2)$ violation is from the 
TeV scale gauge bosons.  The massive $W'$ never gives any $SU(2)_C$ 
violating contributions to the $W^\pm$ and $Z^0$ mass.  The $B'$ 
typically gives an $SU(2)_C$ violating contribution to the electroweak 
gauge boson masses but the additional contributions from the $W^{r\,\pm}$ 
vector bosons largely cancel this.  Summing the various contributions:
\begin{eqnarray}
\delta \rho = 
- \frac{v^2}{64 f^2} \sin^2 2 \theta' 
+ \frac{v^2}{64 f^2} \sin^2 2\beta \sin^2 \phi  
 .
\end{eqnarray}
The second term is a result of the phase in the Higgs vacuum expectation value
that breaks the $SU(2)_C$ and arises because the $W^{r\,\pm}$ 
interactions are not invariant under rephasing of the Higgs.
The phase is generally taken to be $\frac{\pi}{2}$ to have the Higgs states
fall into CP eigenstates.  This is not generic and requires tuning
$m_{12}^2$ to be small. 
Numerically this contribution is:
\begin{eqnarray}
\alpha^{-1} \delta \rho \simeq \frac{1}{8} \sin^2 2\beta \frac{(1 \TeV)^2}{f^2}
\end{eqnarray}
where the $\sin^2 2\theta'$ term has been dropped because it cancels in the
conversion to $\rho_*$ as will be shown in Sec. \ref{Sec: Limits}.
This prefers $\beta$ to be small which is the direction that is 
radiatively driven by the top sector.   For instance at 
$\sin 2 \beta \sim \frac{1}{3}$, this contribution to $\delta \rho$ 
is negligibly small for $f \sim 700$ GeV.   By going to small $\tan \beta$
the mass of the lightest Higgs becomes rather light, for instance, for $\sin2 \beta \simeq \frac{1}{3}$
the mass of the lightest Higgs is bounded by $m_{h^0} \le v$ with most of the parameter
space dominated by $m_{h^0} \le 150$ GeV.

\subsubsection*{Triplet VEV}

Another possible source of $SU(2)_C$ violation is from a triplet
vacuum expectation value.   The form of the plaquette potential
in Eq. \ref{Eq: Pot2} ensures that the tri-linear couplings are 
of the form: 
\begin{eqnarray}
h_1^\dagger \phi^l_H h_2  - h_2^\dagger \phi^l_H h_1.
\end{eqnarray}
There are two equivalent
ways of calculating the effect, either integrating out $\phi^l_H$
to produce higher dimension operators or by calculating
its vacuum expectation value.
The operator appears as:
\begin{eqnarray}
\LL_{u_H u_1 u_2}=  \lambda \cot 2 \vartheta_\lambda f\; i\; \Tr u_H [ u_1,u_2]
\end{eqnarray}
After integrating out $u_H$ the leading derivative interaction is:
\begin{eqnarray}
\LL_{\eff} = 
- \frac{\cos^2 2 \vartheta_\lambda }{16 f^2} 
\Tr D_\mu [u_1,u_2] D^\mu [u_1,u_2] 
\end{eqnarray}
where $D_\mu$ are the Standard Model covariant derivatives.
Expanding this out there is a term that gives a contribution to $\rho$: 
\begin{eqnarray}
\delta \rho = \frac{v^2}{4 f^2} \cos ^2 2 \vartheta_\lambda
\sin^2 2 \beta \sin^2 \phi
\end{eqnarray}
The approximate $\mathbb{Z}_4$ symmetry of the scalar and gauge sectors
that sets $\vartheta_\lambda \simeq \frac{\pi}{4}$ with $\cos 2 \vartheta_\lambda
\sim 10^{-1}$ meaning that this contribution is adequately small.   

One might also worry that the light triplets in $u_{1,2}$ get tadpoles after electroweak
symmetry breaking (through radiatively generated $h^\dag \phi h$ terms), which due to their 
relatively light masses could lead to phenomenologically dangerous triplet vevs.\footnote{We
thank C. Csaki for pointing out that integrating out heavy quarks might 
generate these terms.}
However,
these light scalars are not involved in canceling off the quadratic divergences to the
higgs masses.  Thus these triplets   
can be safely raised to the TeV scale by introducing ``$\Omega$ plaquettes'' as described  
in \cite{Gregoire:2002ra}, where $\Omega = \exp\left(2\pi i\, T^{r3} \right)=\diag(-1,-1,-1,-1,1).$  
These operators suitably suppress the magnitudes of the light triplet vevs and 
do not affect naturalness. 

\subsubsection*{Two Higgs Doublets}

The $\rho$ parameter also receives contributions from integrating
out the Higgs bosons.  It is known that this contribution can
be either positive or negative.  It is positive generically if
the $H^\pm$ states are either lighter or heavier than all the
neutral states, while it is negative if there are neutral Higgs
states lighter and heavier than it.   The Higgs potential of
this theory generically predicts that the charged Higgs is
the heaviest Higgs boson.  There are four parameters of the
Higgs potential: $m_1^2$, $m_2^2$, $b$, and $\lambda$ where
one combination determines $v = 247 \GeV.$   If $\phi \ne \frac{\pi}{2}$
then this analysis becomes much more complicated.
The contribution to $\rho_*$ from vacuum polarisation diagrams is:
\begin{eqnarray}
\nonumber
\delta \rho_* &=& \frac{\alpha}{16 \pi \sin^2\theta_{\text{w}} m^2_{W^\pm}}
\Big(
F(m^2_{A^0},m^2_{H^\pm})\\
\nonumber
&&
\hspace{0.5in} 
+ \sin^2(\alpha -\beta) \big(F(m^2_{H^\pm}, m^2_{h^0}) -F(m^2_{A^0}, m^2_{h^0}) + 
\delta\hat{\rho}_{\text{SM}}(m^2_{H^0})\big)\\
&&
\hspace{0.5in} 
+ \cos^2(\alpha -\beta)
\big(F(m^2_{H^\pm}, m^2_{H^0}) -F(m^2_{A^0}, m^2_{H^0}) + \delta\hat{\rho}_{\text{SM}}(m^2_{h^0})\big)
\Big)
\end{eqnarray}
where
\begin{eqnarray}
F(x,y) &=& \half(x + y) - \frac{xy}{x -y} \log\frac{x}{y}\\
\nonumber
\delta \hat{\rho}_{\text{SM}}(m^2)&=&  F(m^2,m^2_{W^\pm}) - F(m^2,m^2_{Z^0})\\
&&+ \frac{4 m^2 m^2_{W^\pm}}{m^2 - m^2_{W^\pm}} \log \frac{m^2}{m^2_{W^\pm}}
- \frac{4 m^2 m^2_{Z^0}}{m^2 - m^2_{Z^0}} \log \frac{m^2}{m^2_{Z^0}}
\end{eqnarray}
In two Higgs doublet models setting an upper limit on the lightest Higgs mass from
precision electroweak measurements is less precise.
There can be cancellations but it appears as though the $T$ parameter is quadratically
sensitive to the mass of the heaviest Higgs.  
The spectrum of Higgs generated by the Higgs potential keeps the splittings
between the masses of the Higgs bosons constant:
\begin{eqnarray}
\nonumber
m^2_{H^\pm} - m^2_{A^0} = 2 \lambda v^2
\hspace{0.5in}
m^2_{H^\pm} - m^2_{H^0} = m^2_{h^0}
\end{eqnarray} 
with $m^2_{h^0} \le 4 \lambda v^2 \sin^2 2\beta$.
This means that if $\lambda$ is kept small then the $T$ parameter
is insensitive to the overall mass scale of the Higgs. 
With $\alpha - \beta = \frac{\pi}{4}$ the contribution to $\rho_*$
goes as:
\begin{eqnarray}
\nonumber
\alpha^{-1}\,\delta\rho_* &\simeq& \frac{1}{10} 
- \frac{m_{h^0}^2}{(500 \GeV)^2} 
- \frac{1}{4} \frac{m_{h^0}^2}{m^2_{H^\pm}} 
- \frac{1}{30} \log \frac{m^2_{H^\pm}}{(500\GeV)^2}
\hspace{0.3in} \lambda = \half \\ 
&\simeq& \hspace{0.08in} \frac{1}{3} - \frac{m_{h^0}^2}{(500 \GeV)^2} 
- \frac{1}{2} \frac{m_{h^0}^2}{m^2_{H^\pm}} 
- \frac{1}{30} \log \frac{m^2_{H^\pm}}{(500\GeV)^2}
\hspace{0.31in} \lambda = 1 .
\end{eqnarray}
As $\lambda$ becomes larger the contributions to the $T$ parameter
typically become larger, positive and favouring heavier Higgs
with smaller mass splittings to satisfy precision electroweak fits.
Notice that even for $\lambda = \half$ where the contributions to
$\delta \rho_*$ are quite small the mass of the lightest Higgs is only
bounded by $m_{h^0} \le 350 \GeV$.  However the contributions
to $\rho$ from the gauge boson sector prefer a small $\beta$ to keep
the contributions small, thus favouring a light Higgs.

\subsubsection*{Top Partners}

The top partners provide another source of $SU(2)_C$ violating
operators arising from integrating out the partners to the top
quark: $\tilde{u}$ and $\tilde{u}^c$.   Since this is a Dirac
fermion it decouples in a standard fashion as $y_2$ becomes  
large \cite{Collins:1999rz}.   The contribution 
after subtracting off the Standard Model top quark contribution is:
\begin{eqnarray}
\nonumber
\delta \rho_{t'*} &=& 
\frac{N_c\sin^2 \theta_L}{8\pi^2v^2}\left[\sin^2\theta_L F(m_{t'}^2,m_t^2) + 
F(m_{t'}^2,m_b^2) -F(m_t^2,m_b^2) -F(m_{t'}^2,m_t^2) \right] \\
&\simeq& \frac{N_c \sin^2 \theta_L}{16\pi^2v^2}\left[\sin^2\theta_L m_{t'}^2 +
2 \cos^2 \theta_L \frac{m_{t'}^2\:m_t^2}{m_{t'}^2-m_t^2}\log \frac{m_{t'}^2}{m_t^2}
-(2-\sin^2 \theta_L )m_t^2\right]
\end{eqnarray}
where $\theta_L$ is the $t'$ and $t$ mixing angle after electroweak symmetry breaking
and can be expressed in terms of the original Yukawa and the mixing angle $\vartheta_y$: 
\begin{eqnarray}
\sin \theta_L \simeq  \frac{v \sin^2 \vartheta_y \cos \beta}{2 f}
\end{eqnarray}
Using this and the expressions for the mass of the $t$ and $t'$
in Eq. \ref{Eq: Top Mass} the expression for the $\delta\rho_{t'*}$
parameter reduces to:
\begin{eqnarray}
\delta \rho_{t'*} &\simeq& 
\frac{3 y_{\text{top}}^2  v^2 \sin^4 \vartheta_y \cos^4 \beta}{128 \pi^2 f^2}
\Big(
\tan^2 \vartheta_y  -2 
\big( \log\frac{v^2 \sin^2 \vartheta_y \cos^2 \vartheta_y \cos^2\beta}{4 f^2}
+ 1\big) \Big) 
\end{eqnarray}
This contribution vanishes as $\vartheta_y \rightarrow 0$
which is the limit $y_1 \rightarrow 0$ while keeping $y_{\text{top}}$
fixed.  In the limit of $\vartheta_y = \frac{\pi}{4} - \delta \vartheta_y$ 
near where $m_{t'}$ is minimised, the contribution for small $\beta$ goes as:
\begin{eqnarray}
\alpha^{-1}\delta\rho_{t'*} \simeq \frac{ (1 - 4.4\, \delta \vartheta_y+7.5\,\delta\vartheta_y^2)}{25}
\Big( 1 - 1.8 \sin^2 \beta  +0.7 \sin^4 \beta \Big) \frac{(1 \TeV)^2}{f^2}.
\end{eqnarray}
This is adequately small for any $\beta$ and the contribution quickly drops with
$\delta \vartheta_y$.  For instance, with $\delta \vartheta_y \simeq 0.1$, 
$\delta\rho_{t'*}$ drops by 40\% while $m_{t'}$ only rises by 2\%.
This means that this contribution can be taken to be a subdominant effect.

\subsection{$S$ parameter}

The main source for contributions to the $S$ parameter is
from integrating out the physical Higgs bosons.  As for
the case with the $\rho$ parameter, a two Higgs doublet
spectrum leaves a great deal of room for even a heavy 
spectrum where all the states are above 200 GeV.  Generically
the $S$ parameter does not lead to any constraints in the Higgs
spectrum because of cancellations:
\begin{eqnarray}
\nonumber
S &=& \frac{1}{12 \pi}\Big( 
\sin^2(\beta -\alpha) \log\frac{m^2_{H^0}}{m^2_{h^0}}
- \frac{11}{6} + \\
&&
\hspace{0.3in}
\cos^2(\beta -\alpha) G(m^2_{H^0}, m^2_{A^0}, m^2_{H^\pm})
+\sin^2(\beta -\alpha) G(m^2_{h^0}, m^2_{A^0}, m^2_{H^\pm})\Big)
\end{eqnarray}
where
\begin{eqnarray}
G(x,y,z) = \frac{x^2 + y^2}{(x - y)^2}
+ \frac{(x- 3 y)x^2 \log\frac{x}{z} - (y - 3x) y^2 \log \frac{y}{z}}{(x-y)^3} .
\end{eqnarray}
This can be approximated by expanding around large $m^2_{H^\pm}$ masses and taking 
$\alpha - \beta = \frac{\pi}{4}$:
\begin{eqnarray}
S = S_{\text{SM}}-
\frac{5}{144 \pi} - \frac{1}{16\pi} \frac{2 \lambda v^2}{m^2_{H^\pm}}
+ \frac{1}{48\pi} \frac{m^2_{h^0}}{m^2_{H^\pm}} 
+ \frac{1}{24\pi}\log \frac{m^2_{H^\pm}}{m^2_{h^0}}
\end{eqnarray}
These are adequately small in general for all reasonable values of
$\lambda$ and $m^2_{h^0}$.

\subsection{Electroweak Currents}

The last source of electroweak constraints comes from
the modifications to electroweak currents and 
four Fermi operators at low energies.  
These come from two primary sources, the Higgs-Fermion interactions
from the current interactions in Eqs. \ref{Eq: Higgs Currents} and
\ref{Eq: Fermion Currents}:
\begin{eqnarray}
\nonumber
\LL_{\text{H F}} &=& 
-\frac{j_\mu^a{}_{W' \text{H}}\; j^\mu{}^a{}_{W' \text{F}}}{M^2_{W'}}
-
\frac{j_\mu{}_{B' \text{H}}\; j^\mu{}_{B'\text{F}}}{M^2_{B'}}\\
&=&-   \frac{\sin^2 \theta \cos 2\theta}{8 f^2} 
j_{\text{H}}{}^{a\,\mu} j_{\text{F}}{}_{a\,\mu} 
-   \frac{\sin^2 \theta' \cos 2\theta'}{8 f^2} 
j_{\text{H}}{}^{\mu} j_{\text{F}}{}_{\mu} 
\end{eqnarray}
and the direct four Fermi interactions:
\begin{eqnarray}
\nonumber
\LL_{\text{F F}} &=&
- \frac{(j_\mu^a{}_{W'\text{F}})^2}{2 M^2_{W'}}
- \frac{(j_\mu{}_{B'\text{F}})^2}{ 2 M^2_{B'}}\\
&=&
- \frac{\sin^4\theta}{8 f^2}
j_{\text{F}}{}^{a\,\mu} j_{\text{F}}{}_{a\,\mu}
-\frac{\sin^4\theta'}{8 f^2}
j_{\text{F}}{}^{\mu} j_{\text{F}}{}_{\mu} .
\end{eqnarray}

It requires a full fit to know what the limits on these interactions are,
but to first approximation these interactions are fine if they are
suppressed by roughly $\Lambda_{\text{lim}}\sim 6 \TeV$\cite{Barbieri:2000gf}.   Since 
$\sin \theta \simeq \sqrt{3} \sin \theta'$, the biggest constraints come 
from the effects of the $W'$. The constraints reduce to a limit on
the $g_5 - f$ plane of:
\begin{eqnarray}
\frac{2\sqrt{2}f}{\sin \theta} \gsim \Lambda_{\text{lim}} .
\end{eqnarray}
Clearly for $f \sim 2.5$ TeV there are no limits on $g_5$, for
$f \sim 1.5$ TeV, $g_5 \sim 1.5$ and for $f\sim 0.7$ TeV, $g_5\sim 3$. {}\footnote{
It is not possible to push $g_5$ much larger than $3$ because perturbativity
is lost when the loop factor suppression $T_2(A) g_5^2/8 \pi^2$ becomes
roughly 1.   This requires $g_5 \lsim 5$.}
These are clearly all in the natural regime for the little Higgs mechanism 
to be stabilising the weak scale.  This limit is very closely related to the
mass of the $W'$:
\begin{eqnarray}
M_{W'} \gsim \frac{g}{\sqrt{2} \cos\theta} \Lambda_{\text{lim}}
\end{eqnarray}
Thus, the mass of the $W'\gsim \frac{2}{5} \Lambda_{\text{lim}}$.
This sets a lower limit on the mass of the $W'$ of $2.5$ TeV.

\subsection{Summary of Limits}
\label{Sec: Limits}

To state the limits it is necessary to convert $\rho$ to $\rho_*$
which is related to the $T$ parameter.   While $\rho$ is
related to custodial $SU(2)$, $\rho_*$ is related to 
physical results and differs from $\rho$ when there
are modifications to electroweak current interactions.  The difference is due
to the discrepancy between the pole mass of the
$W^\pm$ and the way that the mass of the $W^\pm$ is extracted
through muon decay. 

In this model the Standard Model fermions couple to the $W'$ and $B'$
and integrating out the heavy gauge bosons generates both four Fermi 
interactions and corrections to the $J_Y,J_W$ fermionic currents after electroweak
symmetry breaking.  
Following the analysis in \cite{Csaki:2002qg,Burgess:1993vc}, 
the Fermi constant is corrected by:
\begin{eqnarray}
\frac{1}{G_F}=\sqrt{2}v^2\left(1+\frac{\delta M_W^2}{M_{W_0}^2}
-\frac{v^2}{64f^2}\sin^2 2\theta\right).
\end{eqnarray}
To determine $\rho_*$, it is necessary to integrate out the $Z^0$ and 
express the four Fermi operators as
\begin{eqnarray}
-\frac{4G_F}{\sqrt{2}}\rho_*(J_3-s_*^2J_Q)^2 + \alpha J_Q^2
\end{eqnarray}
which gives us to order $(v^2/f^2)$
\begin{eqnarray}
\nonumber
\delta \rho_* &=& \alpha T = \frac{\delta M_W^2}{M_{W_0}^2} -
 \frac{\delta M_Z^2}{M_{Z_0}^2} + \frac{v^2}{64f^2}\sin^2 2\theta'\\
&=& \delta \rho + \frac{v^2}{64f^2}\sin^2 2\theta'.
\end{eqnarray}
Because all the other contributions to $\rho$ are small,
the primary limit on the theory comes from the $SU(2)_C$ violation
in the gauge sector. 

At this point the limits can be summarised for the masses of the particles. 
The limit on the breaking scale, $f$, is roughly 700 GeV from the contributions to $T$
from the gauge bosons.   The Higgs contributions to $\rho_*$ could have been large,
but because $\tan\beta$ is small it turns out to be subdominant.
The mass of the lightest Higgs is bounded to be less than 250 GeV with most
of the parameter space dominated by masses less than 150 GeV.
The TeV scale vector bosons are all roughly degenerate with masses greater
than 2.5 TeV.  The mass of the top partner is roughly 2 TeV.  While
the mass of the heavy Higgs are roughly 2 TeV from the limits on $f$.

If we chose to exclude the $A^{FB}_b$ measurement as an outlier, the implications for this model 
are significant.  Discarding this measurement might be reasonable since
it deviates from other Standard Model measurements by roughly $3\sigma$.
This model does not significantly alter the physics of $A^{FB}_b$ from the Standard Model.
This measurement is not generally excluded  because 
doing so pulls the fit for the $T$ parameter positive which favours a very light 
Higgs in the Standard Model and is excluded by direct searches.  
However there are additional positive contributions
that mimic a light Higgs boson in this model.  On a general principle, the connection between
a light Higgs boson and a positive contribution to the $T$ parameter
does not hold in two Higgs doublet models and it is quite easy to have the
Higgs sector produce $\delta T \sim 0.2$.    By ignoring $A^{FB}_b$ the
best fit for the $S-T$ plane moves to $T \sim 0.15\pm 0.1$.
See \cite{Chanowitz:2001bv,Chanowitz:2002cd} for more details.
This significantly reduces the constraints on this model because all 
TeV scale physics pulls towards positive $T$.  The contribution from
the gauge bosons becomes roughly about the best fit for $T$ even with 
$\tan \beta \sim 1$ and $f \sim 700\GeV$.  This in turn can lower the limit 
on $m_{t'}$ and also remove the preference for lighter Higgs.

\section{Conclusions and Outlook}
\setcounter{equation}{0}
\renewcommand{\theequation}{\thesection.\arabic{equation}}
\label{Sec: Conclusions}

In this paper we have found a little Higgs model with custodial
$SU(2)$ symmetry that is easily seen to be consistent with precision
electroweak constraints.
This demonstrated that little Higgs models are viable models 
of TeV scale physics that stabilise the weak scale and that
the breaking scale, $f$, can be as low as $700 \GeV$ without being in
contradiction to precision electroweak observables.
This theory is a small modification to the Minimal Moose 
having global $SO(5)$ symmetries in comparison  to $SU(3)$.
Most of the qualitative features of the Minimal Moose carried over
into this model including that it is a two Higgs doublet model with
a coloured Dirac fermion at the TeV scale that cancels the one loop
quadratic divergence of the top and several TeV scale vector bosons.
By having custodial $SU(2)$ it is possible to take the
simple limit where the $g_5$ coupling is large where the
contributions from TeV scale physics to precision electroweak observables
become small.  In the model presented, a breaking scale as low as 
$f = 700 \GeV$ was allowed by precision electroweak observables. 
The limits on the $W'$ and $B'$ are around $2.5 \TeV$ and the mass of the
top partner is roughly $2 \TeV$.  
These are the states that cancel the one loop quadratic divergences from
the Standard Model's gauge and top sectors and their masses are where
naturalness dictates.
The charged Higgs boson was typically the heaviest amongst the light
Higgs scalars this resulted in a positive contribution to $T$.
The limits from custodial $SU(2)$ violating operators favoured a 
light Higgs boson coming not from the standard oblique corrections
from the Higgs boson, but indirectly from integrating out the TeV
scale gauge bosons.  These already mild limits might be reduced by
going away from a maximal phase.  Changing this phase would also require 
recalculating the contributions to $\delta \rho$ from the Higgs sector 
when the states 
do not fall into CP eigenstates.  There are additional scalars that could
be as light as 100 GeV that came as the $SO(5)$ partners to the Higgs.
As mentioned earlier in the section on triplet vevs, these states can be lifted by 
``$\Omega$ plaquettes'' to the multi-TeV
scale and therefore their relevance for phenomenology is model dependent. 

This model predicts generically a positive contribution to $T$ mimicking
the effect of a light Higgs in the Standard Model.  
This is interesting because if one excludes the $A^{FB}_b$ measurement as an
outlier then the fit to precision electroweak observables favours
a positive $T \sim 0.15 \pm 0.1$.  This is generally stated as the Standard Model
has a best fit for a Higgs mass of $40 \GeV$ if the $A^{FB}_b$ measurement
is excluded. 

There has been recent interest in the phenomenology of the Higgs
bosons inside little Higgs models.  Most of the recent work we believe
carries over qualitatively including the suppression of 
$h\rightarrow gg, \gamma\gamma$ \cite{Dib:2003zj,Han:2003gf}.
The LHC should be able to produce copious numbers of the TeV scale
partners in the top and vector sectors \cite{Burdman:2002ns}.  

Another possible way of removing limits arising from the phase in the
Higgs vacuum expectation value is to construct a model that has only one 
Higgs doublet.  All ``theory space'' models automatically have two Higgs 
doublets so one possibility would be to follow the example of the 
``littlest Higgs'' and construct a coset model such as $SO(9)/(SO(5)\times SO(4))$ 
\cite{Chang:To Appear}. 
There may be other two Higgs doublet models that have a gauged $SU(2)_r$ 
that do not force the Higgs vacuum expectation value to break $SU(2)_C$.

To summarise the larger context of this model, it provides a simple realistic
little Higgs theory that is parametrically safe from precision electroweak measurements. 
While it is not necessary to have a gauged $SU(2)_r$, it allows for transparent
limits to be taken where the TeV scale physics decouples from the physics 
causing constraints while still cutting off the low energy quadratic divergences. 
There are other ways of avoiding large contributions to electroweak precision observables
without a gauged $SU(2)_r$.
The important issue is that the physics that is stabilising the weak scale from the 
most important interactions is not providing significant constraints on little Higgs models.
This is the deeper reason why the model presented worked in such a simple fashion.
Precision electroweak constraints are coming from the interactions of either
the $B'$ or the interactions of the light fermions.  The quadratic divergence
from $U(1)_Y$ only becomes relevant at a scale of 10 -- 15 TeV and is oftentimes 
above the scale of strong coupling for little Higgs models.  The interactions
of the light fermions with the TeV scale vector bosons is not determined by
electroweak gauge symmetry and can be altered by either changing the charge assignments
or by mixing the fermions with multi-TeV scale Dirac fermions.

In a broader view little Higgs models offer a rich set of models
for TeV scale physics that stabilise the weak scale.  Each little Higgs model
has slightly different contributions to precision electroweak observables, but
they do not have parametric problems fitting current experimental measurements.   
In the next five years the LHC will provide direct probes of TeV scale physics
and determine whether little Higgs models play a role in stabilising the weak scale.

\section*{Acknowledgments}

We would like to thank N. Arkani-Hamed, T. Gregoire, 
C. Kilic, R. Mahbubani, and M. Schmaltz for useful discussions and comments on this work.
We would also like to thank C. Csaki for pointing out that  
radiatively generated tadpoles could give additional triplet vevs.
S.C. is supported through an NSF graduate student fellowship. 

\appendix
\section{Generators}
\setcounter{equation}{0}
\renewcommand{\theequation}{\thesection.\arabic{equation}}
\label{App: Generators}

The $SO(5)$ commutation relations are:
\begin{eqnarray}
[T^{mn},T^{op}] = \frac{i}{\sqrt{2}} ( 
\delta^{mo} T^{np}
-\delta^{mp} T^{no}
-\delta^{no} T^{mp}
+\delta^{np} T^{mo})
\end{eqnarray} 
where $m,n,o,p$ run from $1,\ldots, 5$.  These generators can be
broken up into
\begin{eqnarray}
\nonumber
T^{l\,a} = \frac{1}{2\sqrt{2}}\epsilon^{abc} T^{bc} + \frac{1}{\sqrt{2}}T^{a4}
\hspace{0.5in}
T^{r\,a} = \frac{1}{2\sqrt{2}} \epsilon^{abc} T^{bc} - \frac{1}{\sqrt{2}}T^{a4}\\
T^{v\,0} = T^{45}
\hspace{0.5in}
T^{v\,a} = T^{a5}
\end{eqnarray}
The commutation relations in this basis are of $SO(5)$ are
\begin{eqnarray}
\nonumber
[T^{l\,a}, T^{l\,b}] = i \epsilon^{abc} T^{l\,c}, 
\hspace{0.5in}
[T^{r\,a},T^{r\,b}] = i \epsilon^{abc} T^{r\,c}, 
\hspace{0.5in}
[T^{l\,a}, T^{r\,b}] = 0, \\
\nonumber
[T^{v\,0},T^{l\,a}] = -[T^{v\,0},T^{r\,a}] = \frac{i}{2} T^{v\,a}, 
\hspace{0.5in}
[T^{v\,0},T^{v\,a}] = \frac{i}{2} (T^{r\,a}-T^{l\,a}), \\
\nonumber
[T^{v\,a},T^{l\,b}] = -\frac{i}{2} T^{v\,0} \delta^{ab} 
                       + \frac{i}{2} \epsilon^{abc} T^{v\,c}, 
\hspace{0.5in}
[T^{v\,a},T^{r\,b}] = \frac{i}{2} T^{v\,0} \delta^{ab} 
                       +\frac{i}{2} \epsilon^{abc} T^{v\,c},
\\
\left[ T^{v\,a},T^{v\,b} \right] = \frac{i}{2} \epsilon^{abc} (T^{l\,c} +T^{r\,c}).
\hspace{1.5in}
\end{eqnarray}

\subsubsection*{Vector Representation}

The vector representation of $SO(5)$ can be realised as:
\begin{eqnarray}
T^{mn}{\,}^{op} = \frac{-i}{\sqrt{2}}( \delta^{mo} \delta^{np} - \delta^{no} \delta^{mp})
\end{eqnarray}
where $m,n,o,p$ again run over $1, \ldots, 5$ and $m,n$ label the 
$SO(5)$ generator while $o,p$ are the indices of the vector representation.
In this representation:
\begin{eqnarray}
\Tr\; T^A T^B = \delta^{AB} .
\end{eqnarray}

\subsubsection*{Spinor Representation}

The spinor representation is given by the form
\begin{eqnarray}
\nonumber
\sigma^{l\,a} = \left( \begin{array}{cc}
\sigma^a/2 & 0 \\
0 & 0 \\
\end{array}\right)
\hspace{0.5in}
\sigma^{r\,a} = \left( \begin{array}{cc}
0 & 0 \\
0 & \sigma^a/2 \\
\end{array}\right), 
\\
\sigma^{v\,0} = \frac{1}{2\sqrt{2}}\left( \begin{array}{cc}
0 & \identity \\
\identity & 0 \\
\end{array}\right) 
\hspace{0.5in}
\sigma^{v\,a} = \frac{1}{2\sqrt{2}}\left( \begin{array}{cc}
0 & i\sigma^a \\
-i\sigma^a & 0 \\
\end{array}\right)
\end{eqnarray}
In this representation
\begin{eqnarray}
\Tr\; T^A T^B  = \frac{1}{2} \delta^{AB}.
\end{eqnarray}

\end{document}